\documentclass[twocolumn,secnumarabic,footinbib,tightenlines,nobibnotes,aps,unsortedaddress,prl]{revtex4-1}
\usepackage{amsfonts}
\usepackage{mathtools}
\usepackage{hyperref}
\usepackage{grffile}

\usepackage[caption=false]{subfig}
\usepackage{graphicx,xcolor}
\usepackage{mathtools}
\usepackage{paralist}
\begin{document}

\newcommand{\nd}{^{\vphantom{\dagger}}}
\newcommand{\dg}{^{\dagger}}
\renewcommand{\vec}[1]{{\boldsymbol{#1}}}
\newcommand{\op}{\hat}
\newcommand{\uvec}[1]{{\vec{e}_{#1}}}
\def\Bk{{\vec k}}
\newcommand{\id}{\mathbb{I}}
\newcommand{\wt}{\widetilde}
\newcommand{\Tr}{\textsf{Tr}}
\newcommand{\Spec}{\textsf{Spec}}
\newcommand{\Det}{\textsf{det}}
\newcommand{\Arg}{\textsf{arg}}
\newcommand{\Sgn}{\textsf{sgn}}
\newcommand{\pr}{^{\prime}}
\def\frac#1#2{{\textstyle{#1 \over #2}}}
\def\half{\frac{1}{2}}
\newcommand{\smat}[1]{\left(
    \begin{smallmatrix}
      #1
    \end{smallmatrix}
\right)}
\def\vk{{\vec{k}}}
\newcommand{\header}[1]{\paragraph{#1}} 
\title{Topological indices for open and thermal systems via Uhlmann's phase}

\author{Zhoushen Huang}
\author{Daniel P. Arovas} \affiliation{Department of Physics,
  University of California, San Diego, CA 92093, USA}

\date{\today}
\begin{abstract}
  Two-dimensional topological phases are characterized by TKNN
  integers, which classify Bloch energy bands or groups of Bloch
  bands. However, quantization does not survive thermal averaging or
  dephasing to mixed states. We show that using Uhlmann's parallel
  transport for density matrices ({\sl Rep.~Math.~Phys.~}{\bf 24}, 229
  (1986)), an integer classification of topological phases can be
  defined for a finite generalized temperature $T$ or dephasing
  Lindbladian. This scheme reduces to the familiar TKNN classification
  for $T<T_{{\rm c},1}$, becomes trivial for $T>T_{{\rm c},2}$, and
  exhibits a `gapless' intermediate regime where topological indices
  are not well-defined. We demonstrate these ideas in detail, applying
  them to Haldane's honeycomb lattice model and the
  Bernevig-Hughes-Zhang model, and we comment on their generalization
  to multi-band Chern insulators.
\end{abstract}
\maketitle

\paragraph{Introduction:}
The discovery of integer quantum Hall effect and its subsequent
theoretical formulation heralded a new paradigm of thinking in
condensed matter physics, which has by now blossomed into the rapidly
growing field of topological phases \cite{BernevigHughes,Hasan10-rmp,Qi11-rmp}.
In integer quantum Hall systems, the Hall conductance $\sigma_{xy}$ is an integer
multiple $C$ of $e^2/h$, where $C$ is the first Chern index for the projector onto
the filled Bloch bands of the system, as first pointed out in a seminal paper by
TKNN \cite{TKNN}. Since an integer cannot change continuously, $\sigma_{xy}$
is robust against perturbations to the system as long as the bulk energy gap is finite, and is said to be
topologically protected.  In symmetry-protected topological (SPT) systems, the Chern number itself may vanish by symmetry,
but one can still define a topological index, using a restricted set of wave functions (\emph{e.g.} a subset of bands \cite{Sheng06,Prodan09},
or within a subspace in the Brillouin zone \cite{Fu06}, \emph{etc.}), which remains a nonzero integer,
an example being the spin Chern number in quantum spin Hall systems.
The Chern number is thus of central importance in the topological characterization of two
dimensional (2D) band insulators.  Another example is that of
topological Floquet systems \cite{LKFA90,Kitagawa10,Lindner11,Gu11}, which
generally are open systems coupled periodically to an environment, typically idealized
as pure eigenstates of the time evolution operator over one period.

Topological classifications and topological phase transitions thus far
have been defined for systems at $T=0$. When dealing with mixed
states, such as $T>0$, quantization is lost. For example, the thermal
average of the Chern number would include a contribution from bands
not filled at zero temperature, and would no longer be an integer. Any
extension of discrete phase classifications to mixed states should be
elicited by density matrices \cite{Viyuela2012,Rivas2013}. Two natural desiderata of such a scheme
would be that it reproduce the familiar TKNN or ${\mathbb Z}_2$
classification for pure states, and that it be topologically trivial
for $T\to\infty$.  

An approach to marrying dephasing in open systems with quantized
response functions such as $\sigma_{xy}$ was developed recently by
Avron and collaborators \cite{Avron2011b}, who introduced a notion of
compatibility between dissipative and nondissipative evolution of the
density matrix in a Lindbladian setting, under which the
non-dissipative response of the inverse matrix of response
coefficients is immune to dephasing. The compatibility condition is
highly nongeneric, and below we shall show how dephasing can still
allow for a discrete classification of topological phases in a more
general setting. Our approach is based on Uhlmann's definition of
parallel transport for density matrices \cite{Uhlmann86,Hubner93},
which maps a cyclic path in a space of density matrices to a matrix
$M$ (see Eqn. \ref{eq:M} below). The simplest prescription is to
examine Uhlmann's phase, $\wt \gamma\equiv \Arg[\Tr\,M]$, which has
been studied in the context of quantum information, and may be
experimentally measurable \cite{Ericsson03,Aberg07,RZ06,Zhu11-NMR}.
Recently Viyuela \emph{et al.}~\cite{Viyuela14} used $\wt \gamma$ to
identify topological transitions in 1D fermion systems at finite
temperature, where $\wt \gamma$ changes discretely from $\pi$ to $0$
at a critical temperature $T_{\rm c}$. For 2D systems, which are the
focus of this work, we will compute $M$ at each $k_x$, and study the
spectral flow of its (complex) eigenvalues with respect to $k_x$. This
will be demonstrated in detail using Haldane's honeycomb lattice model
\cite{Haldane88} and the Bernevig-Hughes-Zhang model \cite{BHZ06}. We
will briefly comment on its application to more general multi-band
Chern insulators such as the Hofstadter model \cite{hof76}.

\paragraph{Parallel transport and geometric phases of open systems:}
The geometric content over a cyclic path of density matrices can be
understood using Uhlmann's parallel transport \cite{Uhlmann86}; see
also the Supplementary Material (SM). Consider a cyclic path of density matrices,
$\rho_0, \rho_1, \rho_2, \ldots, \rho_N$, with $\rho_a \equiv
\rho(\vec g_a)$ where $\vec g_a$ is some parameter vector, \emph{e.g.}
a Bloch momentum, and $\vec g_N = \vec g_0$. Introduce for each
$\rho_a$ two matrices $W_a$ and $U_a$, where $W_a$ is the
\emph{amplitude} and $U_a$ is unitary, with $W_a=\sqrt{\rho_a}\,U_a$,
hence $\rho_a=W\nd_a\,W\dg_a$. The matrices
$\rho_a, W_a$ and $U_a$ are all square and of equal rank.
Uhlmann's parallel transport is a protocol for determining the $U_a$.
Two amplitudes $W_a$ and $W_b$ are defined as
parallel if the choice of $U_a$ and $U_b$ renders $W_a\dg W\nd_b$ a
non-negative definite hermitian matrix.  This is equivalent to minimizing
the norm $\|W\nd_a-W\nd_b\|$, where $\|A\|=\textsf{Tr}(A\dg A)$.
Uhlmann's condition is the analogue of Pancharatnam's parallelity $\langle \psi_a |
\psi_b\rangle > 0$ for pure states \cite{Pancharatnam56}. The geometric
content is contained in $W\nd_0 W_N\dg$, which is the mismatch of $W_0$ with its
parallel transported version $W_N$.  This is analogous to
the situation {\it vis-a-vis\/} pure states, where the Berry phase is encoded
in the mismatch between a state before and after a parallel transport,
$\exp(i\gamma\nd_{\textsf{Berry}}) =
\Tr\big(|\psi_0\rangle\langle\psi_N\dg|\big)$ \cite{Berry84,WZ84}. Since we will only be
interested in the eigenvalues of $W\nd_0 W_N\dg$, it is convenient to
introduce `holonomy matrix' $M$ which has the same eigenvalues as $W\nd_0 W_N\dg$\,,
\begin{gather}
\label{eq:M}
M \equiv \rho\nd_0\, U\nd_0\, U_N\dg = \rho\nd_0\, U\nd_{01}\, U\nd_{12} \cdots U\nd_{N-1\,N}\ ,
\end{gather}
where $U_{ab} \equiv U\nd_a U_b\dg$. The matrix $M$ is the central
object to be considered in the rest of this paper.

To compute $U_{ab}$, we note that parallelity
allows the polar decomposition $\sqrt{\rho_a} \sqrt{\rho_b} = F_{ab}\,
U_{ab}$, where $F_{ab} = \big(\sqrt{\rho_a}\,\rho_b\, \sqrt{\rho_a}\big)^{1/2}$ is
known as the fidelity. Invoking singular value decomposition yields
\begin{gather}
  \label{eq:Uab}
  \sqrt{\rho_a} \sqrt{\rho_b} = L\nd_a D\nd_{ab} R_b\dg\implies U\nd_{ab} = L\nd_a R_b\dg\ ,
\end{gather}
where $L_a$ and $R_b$ are unitary matrices. $D_{ab}$ is diagonal (note
that $a,b$ do \emph{not} label matrix elements), real, and
non-negative, and consists of eigenvalues of $F_{ab}$. $L_a$, $R_b$
and $D_{ab}$ all have the same matrix size as $\rho_a$ and
$\rho_b$.

The eigenvalues of $M$ are in general complex,
and we write $z\nd_i \equiv r\nd_i \, e^{i\gamma\nd_i}$ in polar form.
We will refer to the phases $\gamma\nd_i$
as the geometric phases of the path that generates $M$. This is
motivated by the fact that at zero temperature, the holonomy $M$
reduces to a Wilson loop operator whose (non-zero) eigenvalues are the
non-Abelian Berry phase factors, see SM.
Note that these geometric phases are not independent due to the
restriction that $\Det (M)$ be real and positive, which follows from
Eqs.~\ref{eq:M} and \ref{eq:Uab}.

\paragraph{Topological characterization of 2D insulators at finite temperature:}
We formulate our scheme for general 2D $N$-band insulators with $L_x
\times L_y$ unit cells, assuming translational invariance and periodic
boundary conditions in both directions. Let $|\psi\nd_{n\vk}\rangle$ be the eigenstates of the momentum space Hamiltonian
$H_{\vec k}$, where $n$ is the band index and $\vec k$ is the Bloch
momentum. At the single particle level, the role of density matrix is
played by the \emph{correlation matrix},
\begin{gather}
  \label{eq:rhok-def}
  \rho\nd_{\vec k} = \sum_{n,n'=1}^N x\nd_{nn'}(\Bk)\,
  |\psi\nd_{n\vk}\rangle\langle\psi\nd_{n'\vk}|
\end{gather}
where $x\nd_{nn'}(\Bk)$ are the density matrix elements at each value of $\Bk$.
For thermal distributions, we take 
$\rho\nd_\Bk = A\nd_\Bk\,e^{-(H\nd_\Bk-\mu N\nd_\Bk)/T}$, where
$\nu\equiv\textsf{Tr}\,\rho\nd_\Bk$ is the number of filled bands at $T=0$, independent of $\Bk$. 
At fixed $k_x$, one can compute the holonomy $M_{k_x}(\mu,T)$ over the
path $k_y \in [ 0, 2\pi]$. Then as $k_x$ sweeps a $2\pi$ cycle, the eigenvalues
$\{z_i\}$ trace closed paths in the complex plane.
A similar picture emerges when one considers Lindbladian evolution,
\begin{equation}
{\dot\rho}=-\frac{i}{\hbar} \big[H\nd_0,\rho\big] +
\sum_j C\nd_j\,\rho\,C\dg_j -\frac{1}{2}\, C\dg_j\, C\nd_j\,\rho -
\frac{1}{2}\,\rho\, C\dg_j\, C\nd_j\ .
\end{equation}
With one Lindblad operator $C=\half\sqrt{\gamma_+}\,\sigma^+ + \half\sqrt{\gamma_-}\,\sigma^-$
connecting two bands, the fixed point of this evolution is $\rho(\infty)=\textsf{diag}(x,1-x)$, where
$x=\gamma\nd_+/(\gamma\nd_+ + \gamma\nd_-)$.  Assuming $\gamma\nd_\pm$ are independent of $\Bk$,
the density matrix is equivalent to a thermal one for a flat band model of the type discussed in Ref. \cite{Viyuela14}.

\begin{figure*}[t]
  \centering
  \subfloat{\includegraphics[width=0.18\textwidth,trim=0 20 4.5 0,clip]{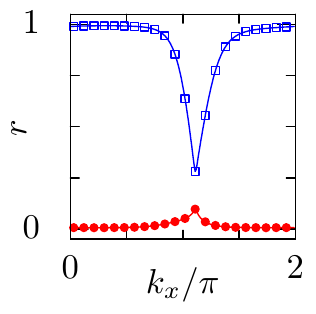}}
  \subfloat{\includegraphics[width=0.155\textwidth, trim=12 20 4.5 0,clip]{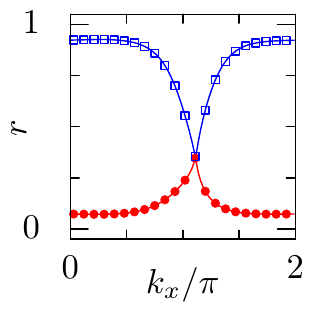}}
  \subfloat{\includegraphics[width=0.155\textwidth, trim=12 20 4.5 0,clip]{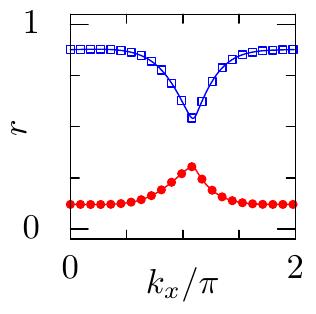}}
  \subfloat{\includegraphics[width=0.155\textwidth,trim=12 20 4.5 0,clip]{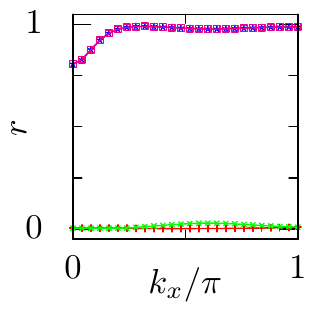}}
  \subfloat{\includegraphics[width=0.155\textwidth, trim=12 20 4.5 0,clip]{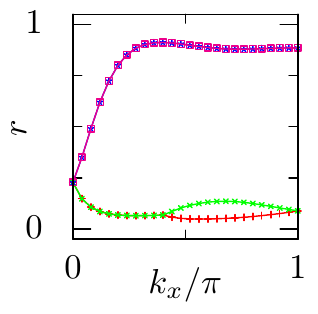}}
  \subfloat{\includegraphics[width=0.195\textwidth, trim=-3 20 0 0,clip]{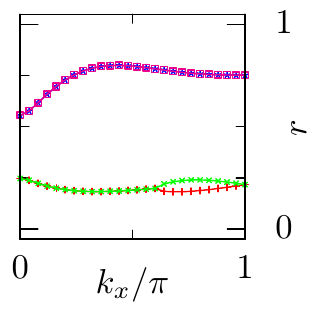}}\\

  \setcounter{subfigure}{0} 
  \vskip -4mm
  \subfloat[$T=0.292$]{\includegraphics[width=0.18\textwidth, trim=3 0 4.5 0,clip]{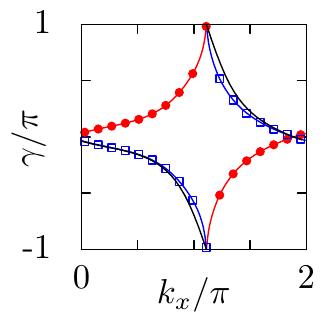}}
  \subfloat[$T=0.6$]{\includegraphics[width=0.155\textwidth,trim=15 0 4.5 0,clip]{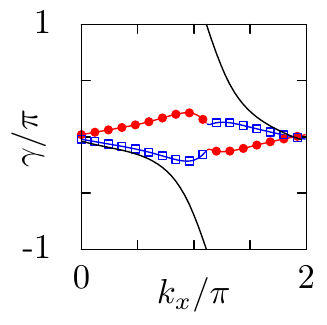}}
  \subfloat[$T=0.8$]{\includegraphics[width=0.155\textwidth, trim=15 0 4.5 0,clip]{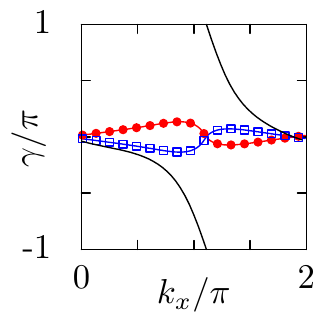}}
  \subfloat[$T=0.2$]{\includegraphics[width=0.155\textwidth, trim=15 0 4.5 0,clip]{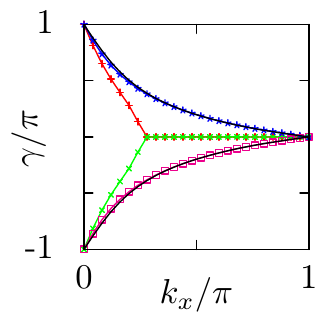}}
  \subfloat[$T=0.4$]{\includegraphics[width=0.155\textwidth,trim=15 0 4.5 0,clip]{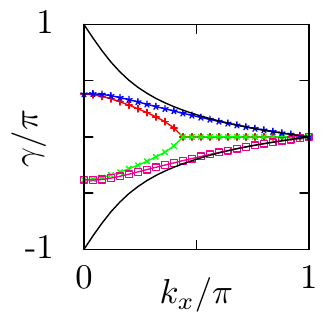}}
  \subfloat[$T=0.7$]{\includegraphics[width=0.195\textwidth, trim=-3 0 3 0,clip]{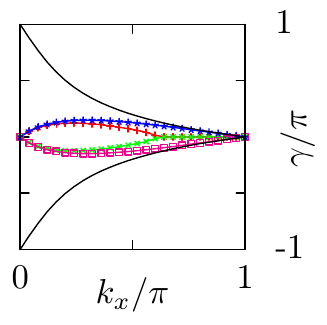}}
  \caption{(Color online) Holonomy eigenvalues of the Haldane model
    (a-c) and the Bernevig-Hughes-Zhang (BHZ) model (d-f). Top panels:
    amplitudes of eigenvalues. Bottom panels: phases of eigenvalues,
    black solid lines correspond to the Berry phase of the lower band
    (a-c) or the non-Abelian Berry phases of the lower two bands
    (d-f), \emph{i.e.}, the $T=0$ limit. Color/point type encodes the
    eigenvalue index. For the Haldane model: in (a), the amplitudes
    are gapped, and the two phases wind in opposite directions. In
    (b), the amplitudes are gapless. The phases do not wind. In (c),
    the amplitudes are gapped again. The phases do not wind. For the
    BHZ model: The amplitudes are gapped in (d) and (f), but gapless
    in (e). In the phases, partner switching occurs in (d) but not in
    (f), see text. Normalization of density matrices is chosen as $\Tr
    \rho_{\vec k} = \nu$ where $\nu$ is the number of filled bands at
    $T=0$. Parameters used for Haldane model: $m=0.5$, $\phi =
    0.3\pi$, $t=0.3$, $\mu = 0.5$. Parameter used for BHZ model:
    $m=1.1$, $\Delta = 0.3$, $\mu=0$. Lattice size: $L_x = 200$, $L_y
    = 50$.}
  \label{fig:haldane-bhz-eig}
\end{figure*}

The topological numbers of the system are to be extracted from the
winding of the $\{z\nd_i(k_x)\}$. However, we need to
take into account that the amplitude spectrum $\{r_i(k_x)\}$ has a gap
structure much like that of Bloch spectra. If a particular level
$r_i(k_x)$ is isolated from the rest, then one can define a winding
number of the corresponding geometric phase, $C_i \equiv
[\gamma\nd_i(2\pi) - \gamma\nd_i(0)]/2\pi$. At zero temperature, since $M$
reduces to the Wilson loop operator, $C_i$ reduces to the winding
number of the $i^{\textsf{th}}$ non-Abelian Berry phase
\cite{ZA12-hof}. For a group of $K$ levels which evolve into each other
but remain isolated from the remainder, the topology of the winding is
naturally characterized as an element of the $K$-string braid group on
the punctured plane, but there are two natural simple choices:
\begin{inparaenum}[(a)]
\item The collective topological number could be the winding number of
  the \emph{sum of the phases}, $\sum_{j=1}^K\Arg(z\nd_j)$. This
  choice is motivated by the analogy with zero temperature gapless
  energy bands, where the total Chern number is the winding number of
  the sum of the individual Berry phases.
\item \label{item:phase-of-sum}Alternatively it could also be defined
  as the winding number of the \emph{phase of the sum}, $\Arg\big(\sum_{j=1}^K z\nd_j\big)$.
  Such a choice draws analogy from multi-path
  interference type experiments, where each complex eigenvalue $z_i$
  encodes both the weight and the phase of the $i^{\textsf{th}}$ path,
  and the output is a coherent sum of these complex weights.
\end{inparaenum}
In both cases, with $g$ spectral gaps in $\{r_i\}$, one obtains $g+1$
topological numbers.
For time reversal invariant topological insulators, one should instead
consider the time-reversal partner switching similar to that of the
non-Abelian Berry phases \cite{Fu06,Yu11,ZSH-unpub}. As temperature
increases from zero, the gap structure of $\{r_i\}$ also changes, and
the system experiences a series of topological transitions until it
reaches a fully trivial stage where all topological numbers are zero.

We note that the spectrum of $M$ in Eqn. \ref{eq:M} depends on the starting
point of the loop. Varying this origin, the general picture of a $T$-dependent
evolution and topological transitions remains unchanged, although the values of
$T$ where transitions occur may vary.  One can then define a critical
temperature by minimizing $T\nd_{\rm c}(k_y^0)$ over the loop origin $k_y^0$.
For some models, as we shall see, an origin-independent transition can be defined.
Additionally, one also can obtain origin-independent transitions in models of
Lindbladian evolution.

\paragraph{Haldane model:}
The Haldane model \cite{Haldane88} describes electrons hopping on a
honeycomb lattice in a fluctuating magnetic field. The model has three
parameters $t, m, \phi$, where $t \equiv
t_{\textsf{NNN}}/t_{\textsf{NN}}$ is the ratio of hopping amplitudes
between second neighbors and first neighbors, $m$ is the Semenoff mass
contrasting the two sublattices, and $\phi$ is a phase associated with
the second neighbor hops which breaks time reversal symmetry.
The ground state can be either a Chern insulator (with $C = \pm 1$) or a trivial
insulator ($C = 0$) depending on the choice of parameters.

In Fig.~\ref{fig:haldane-bhz-eig} (a)--(c), we plot the spectral flow
of holonomy eigenvalues (both amplitudes and phases) as functions of
$k_x$, at three different temperatures. Here $k_x \equiv \vec k \cdot
\vec a_x$ is the Bloch wavevector along the honeycomb basis vector in
the $x$ direction. The matrix $M$ has two eigenvalues for two band
models. From $\Det\, M > 0$, the two geometric phases are opposite to
each other, and we shall focus on $\gamma\nd_>$, the phase
corresponding to the larger eigenvalue magnitude (blue in the figure).

We find that there are three temperature regimes with distinctive
spectral patterns:
\begin{inparaenum}[(i)]
\item In the low temperature regime (panel a), $\gamma\nd_>$ winds
  once, and its spectral flow is a minor deviation from the Berry phase
  flow (\emph{i.e.}~its zero temperature limit). The two amplitudes
  remain gapped. As $T$ increases, the amplitude gap reduces and the
  deviation of $\gamma\nd_>$ increases.
\item In the intermediate temperature regime (panel b), the amplitudes
  touch and stay gapless. There is no winding in the individual
  phases.
\item In the high temperature regime (panel c), the amplitudes are
  gapped again, and $\gamma\nd_>$ does not wind.
\end{inparaenum}

At fixed $k_x$, the correlation
matrices are labeled by the Bloch wavevector along the $\vec a_2$
basis vector of the honeycomb lattice, denoted as $k = \vec k \cdot
\vec a_2$. Since these are $2\times 2$ matrices, we can write
$\sqrt{\rho_k} = \chi_k \Bigl[ \sqrt{1-|\vec b_k|^2} + \vec b_k \cdot
\vec \sigma \Bigr]$, implicitly defining $\chi_k$ and $\vec b_k$
through Eq.~\ref{eq:rhok-def}. In Fig.~\ref{fig:haldane-bhz-eig} (b),
the $k_x$ point at which the amplitude gap closes, denoted as $k_{\rm
  c}$, coincides with where the $\pi$ Berry phase occurs. This is
the \emph{coplanar point}, $k_{\rm c} = \pi+\sin^{-1}(m/6t\sin\phi)$ \cite{ZA12-haldane},
where the entire path of $\vec b_k$ lies on the same plane that passes through the origin.
The winding of $\gamma\nd_>$ over $k_x$ is entirely determined from its
value at $k_x = k_{\rm c}$: it winds once if $\gamma\nd_>(k_{\rm c}) = \pi$,
otherwise it does not wind. The eigenvalues of $M(k_{\rm c})$ are
\begin{gather}
  \label{eq:mpm}
  z\nd_\pm = \frac{1}{2} \Tr(\rho_0) \left[\cos(2S) \pm \sqrt{m^2 -
      \sin^2(2S)}\right]
\end{gather}
with $\rho_0$ the correlation matrix at $(k_x,k_y) = (k_{\rm c},0)$, and
\begin{gather}
  \label{eq:Sz}
  S = \frac{1}{2} \bigg| \oint {\vec b}_k \times
    d{\vec b}_k \bigg|\quad , \quad m = {f\nd_> - f\nd_<\over f\nd_>
    + f\nd_<}\quad ,
\end{gather}
\emph{i.e.}~$S$ is the area enclosed by the path of $\vec b_k$. Here
$f\nd_> \ge f\nd_<$ are the two eigenvalues of $\rho_0$ (\emph{i.e.}
Fermi weights). From Eqs.~\ref{eq:mpm} and \ref{eq:Sz}, we can
understand the spectral evolution as a function of temperature: If
$m < |\sin(2S)|$, $z\nd_\pm$ form a complex conjugate pair, and the
amplitude spectrum is gapless; otherwise they are both real and the
amplitudes are gapped. In the gapped regimes, one can check the zero
and infinite temperature limits: at $T=0$, $S=\pi/2$ \footnote{At
$T=0$, $|\vec b_k| = 1/\sqrt{2}$, which follows from the projective
nature of $\rho_k$ (and hence $\sqrt{\rho_k}$) at $T=0$.} and $m =1$,
yielding $z\nd_+ = 0$ and $z\nd_- = -1$, hence $\gamma\nd_>(T=0) = \pi$,
whereas for $T \rightarrow +\infty$, $m=S=0$, $z\nd_+ = 1$ and $z\nd_- = 0$,
hence $\gamma\nd_>(T\rightarrow \infty) = 0$.

\begin{figure}[t]
  \centering
  \includegraphics[width=0.4\textwidth]{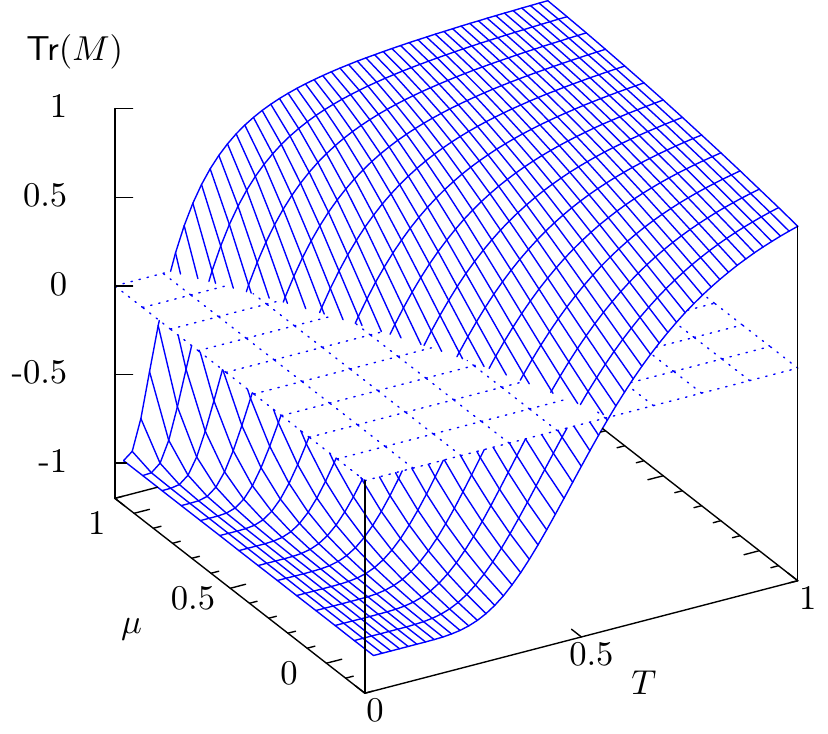}
  \caption{\label{fig:TrM}(Color online) $\Tr\,M$ at $k_x = k_{\rm c}$
    for the Haldane model in the parameter space of temperature $T$
    and chemical potential $\mu$. At $k_x = k_{\rm c}$, $\Tr\,M$ is
    real so $\wt \gamma$ depends solely on its sign. A topological
    transition occurs when $\Tr\,M$ crosses zero, giving $T_{\rm c}$
    as a function of $\mu$. Other parameters are the same as
    Fig.~\ref{fig:haldane-bhz-eig}.}
\end{figure}

As discussed before, in the regime with gapless amplitudes, one has to
consider the winding of a collective phase. Using $\wt \gamma \equiv
\Arg \big[\Tr\, M\big]$ (choice \ref{item:phase-of-sum})
and Eq.~\ref{eq:mpm}, we have, at $k_x = k_{\rm c}$,
$\exp(i\wt \gamma) = \Sgn[\cos(2S)]$.
On the other hand, in the gapped regimes, it follows from $\Det\, M > 0$
that $z\nd_\pm$ and hence $\Tr\, M$ must have the same sign, implying
that $\gamma\nd_> = \wt \gamma$ at $k_{\rm c}$. Thus one can use $\wt
\gamma$ in the entire range of $T$ as \emph{the} geometric phase. The
topological index is entirely determined by $\wt \gamma$ at $k_{\rm c}$, or
equivalently the area $S$ circulated by the path of $\vec b_k$,
regardless of the gap structure of the amplitudes -- a feature of two
band models. Since increasing $T$ generically causes the loop area $S$
to shrink, $\Sgn[\cos(2S)]$ must change from $-$ to $+$ and
not the other way around. One can thus define a unique transition
temperature $T_{\rm c}$ such that $\wt \gamma(k_{\rm c})$ changes discretely
from $\pi$ to $0$. The 2D topological transition coincides with the
effective 1D transition \cite{Viyuela14} at $k_{\rm c}$. Geometrically, such
a topological transition occurs when the loop area $S$ reaches half of
its zero temperature value, $S(T_{\rm c}) = S(0)/2 = \pi/4$.
Note that the same critical temperature $T_c$ would be obtained by
using $M\nd_a = \rho\nd_a U\nd_a U_{a+N}\dg$ for $a \neq 0$, because $S$ is
independent of $a$ according Eqn.~\ref{eq:Sz}.
The transition temperature $T_{\rm c}$ may depend on other external
parameters of the system as well. In Fig.~\ref{fig:TrM}, we allow the
chemical potential $\mu$ to vary, and plot $\Tr\,M = \cos(2S)$ in the
$(\mu,T)$ space. $T_{\rm c}(\mu)$ is determined as the curve at which
$\cos(2S)$ crosses zero.  For Lindbladian evolution, the area is
$S(x)=\frac{\pi}{4}\big(1-2\sqrt{x(1-x)}\big)$, and setting $S(x\nd_{\rm c})=\half\,S(0)$
we obtain $x\nd_{{\rm c},\pm}=\half\pm\frac{\sqrt{3}}{4}$.  Our result for $x\nd_{{\rm c},+}$
corresponds to $T_{\rm c}$ in Ref. \cite{Viyuela14}'s analysis of the flat band case.
We note that $x=\half$ corresponds to $T=\pm\infty$, with $x<\half$ a regime of negative
temperature.

\paragraph{BHZ model:} We briefly discuss the BHZ model \cite{BHZ06}
as an example of the ${\mathbb Z}_2$ class. The Hamiltonian is $H =
\sin k_x \sigma_z \tau_x + \sin k_y \tau_y + (2 - m - \cos k_x - \cos
k_y) \tau_z + \Delta \sigma_y \tau_y$, with $\Delta \neq 0$ breaking
inversion symmetry. The parameters are chosen so that it is in the
topological phase at zero temperature. In
Fig.~\ref{fig:haldane-bhz-eig} (d)--(f), we plot its holonomy spectrum
in the \emph{half}-Brillouin zone. The four amplitudes form two pairs
(partners), which are gapless for intermediate $T$ but gapped for both
low and high $T$. In the low temperature regime (d), each pair of the
geometric phases exhibit partner switching \cite{Fu06,Yu11} where they
change from $\pi$ at $k_x = 0$ to $0$ at $k_x = \pi$, hence the system
is topological. After going through the gapless regime (e), the system
becomes trivial in the high temperature gapped regime (f) where there
is no partner switching.

\paragraph{Conclusion and Discussion:}
In this work, we introduced a topological characterization of 2D band
insulators described by mixed states, resulting from thermal and/or dephasing effects.
The classification is in terms of the winding of geometric phases defined
through Uhlmann's parallel transport of density matrices.  For Haldane's honeycomb
lattice model, we found three phases: (i) a low temperature topological phase classified by
the familiar TKNN integers, (ii) a `gapless' intermediate phase, and (iii) a topologically trivial
high temperature phase.  We found a similar structure in the BHZ model {\it vis-a-vis\/} partner
switching which defined ${\mathbb Z}_2$ quantum numbers.
An analogous procedure works for multi-band Chern insulators as well \cite{ZSH-unpub}, where there is
a series of topological transitions induced by changes in the gap
structure of the amplitude spectrum.

While our primary interest is an extension of topological indices from
pure states to mixed states, we note that it should be experimentally
relevant as well \cite{Ericsson03,Aberg07,Zhu11-NMR,Viyuela14}. In the
context of band insulators, there already exist experimental
techniques to measure Berry phases in 1D \cite{Atala13}
and Chern numbers in 2D \cite{Abanin13}. Such techniques can be extended to measure the Uhlmann phase
$\wt \gamma = \Tr\,M$ through the so-called purification procedure \cite{Ericsson03},
where the amplitude matrix $W$ of a mixed state is mapped to a pure
state in an enlarged system, the reduced density matrix of which is
the mixed state. The Uhlmann phase $\wt \gamma$ of the mixed state is
identified with the Berry phase of the enlarged system, and is thereby
measurable.  The design and implementation of an adiabatic protocol to
measure the Uhlmann phase winding and associated phase transitions
of the types discussed here remains a tantalizing possibility.

\paragraph{Acknowledgements:}
We are grateful to J.~McGreevy, C.~Wu, E. Demler, and Da Wang for discussions.
After this work was completed, we learned of similar results by Viyuela,
Rivas, and Martin-Delgado \cite{VRMunp}.  We thank O. Viyuela for correspondence.
This work was supported by the NSF through Grant No.~DMR-1007028 and
UC Academic Senate.

\bibliographystyle{apsrev-no-url}
\bibliography{uhlmann}

\end{document}


\newcommand{\nd}{^{\vphantom{\dagger}}}
\newcommand{\dg}{^{\dagger}}
\renewcommand{\vec}[1]{{\boldsymbol{#1}}}
\newcommand{\op}{\hat}
\newcommand{\uvec}[1]{{\vec{e}_{#1}}}
\def\Bk{{\vec k}}
\newcommand{\id}{\mathbb{I}}
\newcommand{\wt}{\widetilde}
\newcommand{\Tr}{\textsf{Tr}}
\newcommand{\Spec}{\textsf{Spec}}
\newcommand{\Det}{\textsf{det}}
\newcommand{\Arg}{\textsf{arg}}
\newcommand{\Sgn}{\textsf{sgn}}
\newcommand{\pr}{^{\prime}}
\def\frac#1#2{{\textstyle{#1 \over #2}}}
\def\half{\frac{1}{2}}
\newcommand{\smat}[1]{\left(
    \begin{smallmatrix}
      #1
    \end{smallmatrix}
\right)}
\def\vk{{\vec{k}}}
\newcommand{\header}[1]{\paragraph{#1}} 
\title{Topological indices for open and thermal systems via Uhlmann's phase:\\ Supplementary material }

\author{Zhoushen Huang}
\author{Daniel P. Arovas} \affiliation{Department of Physics,
  University of California, San Diego, CA 92093, USA}

\date{\today}
\maketitle
\onecolumngrid

Here we provide a self-contained review of Uhlmann's parallel
transport following Refs. \cite{Uhlmann86,Hubner93}, and derive its
relation with the Wilson Loop operator at zero temperature, and some
results for two-band models.

Consider a path of density matrices $\rho_1, \rho_2, \ldots, \rho_N$
with $\rho_a \equiv \rho(\vec g_a)$ and $\vec g_a$ is some vector in
the parameter space. We do \emph{not} impose the normalization
$\Tr(\rho_a) = 1$.
Each $\rho_a$ can be \emph{lifted} to a vector $W_a$ in a
Hilbert-Schmidt space,
\begin{gather}
  \label{w-def}
  W_a = \sqrt{\rho_a}\, U_a\ , \ \rho_a = W\nd_a W_a\dg\ ,
\end{gather}
with arbitrary unitary matrix $U_a$. The inner-product and norm in the
Hilbert-Schmidt space are defined as
\begin{gather}
  \big\langle W_a\,,\,W_b\big\rangle = \Tr\,(W_a\dg W\nd_b)\quad , \quad ||w|| =
  +\sqrt{ \langle w \,,\,\ w\rangle}\ .
\end{gather}

\section{Parallelity}
Two vectors $W_a$ and $W_b$ are defined as \emph{parallel} if the
choice of $U_a$ and $U_b$ minimizes $||W_a - W_b||^2$, or
equivalently, $\Tr\,(W_a\dg W\nd_b + W_b\dg W\nd_a)$ is maximal. Polar-decompose
$W_a\dg W\nd_b = \wt HV$ where $\wt H$ is positive-semidefinite hermitian
(denoted as $\wt H \ge 0$) and $V$ is unitary, then
\begin{gather}
  \Tr\,(W_a\dg W\nd_b + h.c.) = \Tr\left[\wt H(V + V\dg)\right]\ .
\end{gather}
The freedom in $U_b$ is equivalent to that of $V$ (assume we fix
$U_a$). Evaluating the trace in the eigenbasis of $V$, one finds that
it is maximized when $V = \id$. Thus parallelity demands
\begin{gather}
  \label{ww-parallel}
  W_a\dg W\nd_b\ge 0\quad \text{(parallelity)}\ .
\end{gather}

Denote the connection between $W_a$ and $W_b$ as
\begin{gather}
  U_{ab} \equiv U_aU_b\dg = (U_{ba})\dg\ .
\end{gather}
Eq.~\ref{ww-parallel} implies that $H \equiv \sqrt{\rho_a}
\sqrt{\rho_b} U_{ba}$ is also hemitian and positive-semidefinite.
Using $H = \sqrt{H H\dg}$, one has the polar decomposition
\begin{gather}
  \label{prod-of-sqrt}
  \sqrt{\rho_a}\,\sqrt{\rho_b} = || \sqrt{\rho_a}\,\sqrt{\rho_b}|| \cdot
  U_{ab} = U_{ab} \cdot || \sqrt{\rho_b}\, \sqrt{\rho_a} ||\ ,
\end{gather}
where the second equality follows from the hermiticity of $U_b (W_a\dg
W_b) U_b\dg$, and the hermitian part is the \emph{fidelity},
\begin{gather}
  \label{amp-def}
  F(\rho_a, \rho_b) \equiv || \sqrt{\rho_a}\,\sqrt{\rho_b}|| = \sqrt{\sqrt{\rho_a}\, \rho_b\, \sqrt{\rho_a}}\ ,
\end{gather}

The connection $U_{ab}$ is invariant under rescaling of the density
matrices $\rho_a \rightarrow \lambda_a\, \rho_a$ where $\lambda_a \in
\mathbb{R}$ can be different for different $a$. In particular,
normalizing $\Tr\, \rho_a = 1$ does not change $U_{ab}$.

If $\rho_{a,b}$ are both invertible, then $U_{ab}$ can be explicitly
written as
\begin{align}
  \notag
  U_{ab} &= || \sqrt{\rho_a}\sqrt{\rho_b}||^{-1}\cdot \sqrt{\rho_a}\,\sqrt{\rho_b} = \sqrt{\rho_a}\,\sqrt{\rho_b} \cdot || \sqrt{\rho_b}
  \,\sqrt{\rho_a}||^{-1}\\
  \label{Uab}
  &= ||\sqrt{\rho_a}\,\sqrt{\rho_b}|| \cdot \sqrt{\rho_a}^{-1}\, \sqrt{\rho_b}^{-1} = \sqrt{\rho_a}^{-1}\, \sqrt{\rho_b}^{-1} || 
  \sqrt{\rho_b}\, \sqrt{\rho_a} ||
\end{align}
the first line follows from eq.~\ref{prod-of-sqrt} and the second line
from $U\nd_{ab} = (U_{ab}^{-1})\dg$

\section{$U_{ab}$ and Singular value decomposition (SVD)}
A practical way to compute the connection $U_{ab}$ is the following.
Standard SVD yields
\begin{gather}
  \sqrt{\rho_a}\, \sqrt{\rho_b} = L\nd_a D\nd_{ab} R_b\dg \quad \text{(no sum)}
\end{gather}
where $L_a$, $R_b$ are unitary matrices and $D_{ab}$ is a diagonal
matrix with non-negative entries,
all three of shape $Q\times Q$. Comparing with Eqs.~\ref{prod-of-sqrt}
and \ref{amp-def} one has
\begin{gather}
  F(\rho_a, \rho_b) = L\nd_a D_{ab} L_a\dg\quad , \quad U_{ab} = L\nd_a R_b\dg\ .
\end{gather}

For simplicity assume all singular values are non-zero ($D_{ab}$ full
rank), then the SVD is unique up to a $Q\times Q$ gauge transformation
under which $D_{ab}$ is invariant,
\begin{gather}
  \label{gauge-trans}
  L_a \rightarrow L_a\, \Lambda\quad , \quad R_b \rightarrow R_b\,
  \Lambda\quad , \quad \Lambda\, D_{ab}\, \Lambda\dg = D_{ab}\ .
\end{gather}
For non-degenerate $D_{ab}$, $\Lambda$ is a diagonal matrix of phase
factors. 
It is possible for several singular values to be zero, \emph{e.g.} at
zero temperature. We shall take a practical stance and assume the
temperature to be infinitesimal instead, which lifts the zero singular
values. See later section on the connection with Wilson Loop for a
treatment of $T=0$.

\section{Parallel transport and the holonomy matrix}
A parallel lift $\{W_0, W_1, W_2, \cdots, W_N\}$ of the density matrices
$\{\rho_0, \rho_1, \rho_2, \cdots, \rho_N\}$ is one where all neighboring $W_a$'s
are parallel. This induces a parallel transport from $U_0$ to $U_N$,
\begin{align}
  U_{0N} &= U_{01}\, U_{12}\, \cdots \,U_{N-1\,N}\\
  \label{lr-series}
  &= L\nd_0\,R_1\dg\, L\nd_1\, R_2\dg \,\cdots\, L\nd_{N-1} R_N\dg \ ,
\end{align}
which depends only on the $\rho$'s following eq.~\ref{Uab}. 
There is a disagreement (i.e. holonomy) in $W$ after a cyclic parallel
transport. Assuming $0$ and $N$ above coincide, we define a holonomy
matrix
\begin{gather}
  M \equiv U\nd_N\, (W_N\dg W\nd_0)\, U_N\dg = \rho\nd_0\, U\nd_{0N}\ .
\end{gather}
As shown in the text, the geometric content of the cyclic path lies in
the spectrum of $M$.

\section{Connection with Wilson Loop and Wannier centers}
Let us now turn to band insulators and treat $\rho_k$ as the one-body
correlation matrix.
At $T = 0$, assume $\nu$
bands are filled out of a total of $Q$ bands. Then $\rho_k$ reduces to
the projector $G_k$,
\begin{gather}
  G\nd_k = \xi\nd_k\,\xi_k\dg\quad , \quad  \xi\nd_k \equiv
  \big\{|\psi\nd_{1k}\rangle\,,\,|\psi\nd_{2k}\rangle\,,\, \ldots \,,\,
  |\psi\nd_{\nu k}\rangle\big\}\ ,
\end{gather}
where $\xi_k$ is a $Q \times \nu$ matrix. In the interest of SVD, we
instead write it as a product of two $Q \times Q$ matrices
\begin{gather}
  G_k = (\xi_k\ , \ 0)
  \begin{pmatrix}
    \xi_k\dg \\ 0
  \end{pmatrix}\ .
\end{gather}
Then the product of two neighboring $\sqrt{\rho_k}$ is ($\sqrt{\rho_k}
= \rho_k = G_k$)
\begin{align}
  G_a G_b &= (\xi_a\ , \ 0)
  \begin{pmatrix}
    \xi_a\dg \xi\nd_b \\ & 0
  \end{pmatrix}
  \begin{pmatrix}
    \xi_b\dg \\ 0
  \end{pmatrix}\\
  &= (\xi_a\ , \ \wt\xi_a)
  \begin{pmatrix}
    \xi_a\dg \xi\nd_b \\ & 0
  \end{pmatrix}
  \begin{pmatrix}
    \xi_b\dg \\ \wt\xi_b\dg
  \end{pmatrix}  \ .
\end{align}
In the second line we have made the first and last matrices unitary by
inserting an arbitrary orthonormal bases for the unoccupied space,
$\wt \xi_k$. These satisfy
\begin{gather}
  \label{g-xi-orth}
  G_k\, \wt \xi_k = 0\ .
\end{gather}
From
\begin{gather}
  \xi_k\dg\, \xi\nd_{k+\delta} = \exp\big[-i\delta A(k) -
  \mathcal{O}(\delta^2)\big]\quad , \quad A\nd_{mn}(k) = \langle \psi\nd_{mk}|
  \,i\,\partial\nd_k\, | \psi\nd_{nk}\rangle\ ,\ (m,n = 1,2, \cdots ,\nu)
\end{gather}
$\xi_a\dg \xi_b$ is unitary for $k_b - k_a \rightarrow 0$, thus
absorbing it to (say) the last matrix, we have
\begin{gather}
  G_aG_b = L_a
  \begin{pmatrix}
    \id \\ & 0
  \end{pmatrix} R_b\dg\quad , \quad L_a = (\xi_a\ , \ \wt \xi_a)\quad
  , \quad R_b\dg =
  \begin{pmatrix}
    \xi_a\dg\, G\nd_b \\ \wt \xi_b\dg
  \end{pmatrix}
\end{gather}
hence the connection is
\begin{gather}
  U\nd_{ab} = L\nd_a R_b\dg = G\nd_a G\nd_b + \wt \xi\nd_a \wt \xi_b\dg\ .
\end{gather}
Parallel transporting forward and using eq.~\ref{g-xi-orth},
\begin{gather}
  U\nd_{ab}\,U\nd_{bc} = G\nd_a\,G\nd_b\,G\nd_c + \wt\xi\nd_a \,\wt \xi_c\dg\ .
\end{gather}
Upon a cyclic path one then has
\begin{gather}
  U_{0N} = G_0\,G_1 \cdots G_N + \wt \xi_0\, \wt \xi_N\dg
\end{gather}
Finally the holonomy matrix is
\begin{gather}
  M = G_0\, U_{0N} = G_0\, G_1 \cdots G_N\ ,
\end{gather}
where the connection in the unoccupied space is projected out by
$G_0$. We thus recover the Wilson loop at zero temperature. The phases
of its eigenvalues are the non-Abelian Berry phases, which are
physically the offsets of Wannier centers from unit cell boundaries.

\section{$2$-band models}
A $2\times 2$ matrix $M$ satisfies $M^2 - M\, \Tr\, M + \Det\, M =
0$.
Thus
\begin{gather}
  M = {M^2 + \Det\, M\over\Tr\, M}\quad , \quad \Tr\, M = \sqrt{\Tr\, (M^2 + \Det M)}\ .
\end{gather}
The second equation is obtained from taking the trace of the first
equation. Now let $M = \sqrt{\sqrt{\rho_1} \,\rho_2\, \sqrt{\rho_1}}$. $M$
is positive-semidefinite. Further denote
\begin{gather}
  \sqrt{\rho_i} = a_i + \wt{\vec b}_i \cdot \vec \sigma \ .
\end{gather}
Using $\sqrt{\rho_i}^{-1} = (a_i - \wt{\vec b}_i\cdot \vec \sigma) /
\Det \sqrt{\rho_i}$, eq.~\ref{Uab} yields
\begin{gather}
  \label{u12-2band}
  U_{12} = {\sqrt{\rho_1} \,\sqrt{\rho_2} + \Det\, \sqrt{\rho_1}\, \Det\,
    \sqrt{\rho_2}\, \sqrt{\rho_1}^{-1} \sqrt{\rho_2}^{-1}
  \over\sqrt{\Tr\left(\rho_1\rho_2 + \Det \sqrt{\rho_1}\Det \sqrt{\rho_2}
      \right)}} = {a_1 a_2 + \wt{\vec b}_1 \cdot \wt{\vec b}_2 + i (\wt{\vec
    b}_1 \times \wt{\vec b}_2) \cdot \vec \sigma\over\sqrt{(a_1a_2 + \wt{\vec b}_1
      \cdot \wt{\vec b}_2)^2 + (\wt{\vec b}_1\times \wt{\vec b}_2)^2}}\ .
\end{gather}
The RHS is explicitly unitary. One can verify that
\begin{gather}
  U_{k, k+\delta k} = 1 + i \delta k\,{\wt{\vec b}_k \times \partial_k\,
    \wt{\vec b}_k\over a_k^2 + |\wt{\vec b}_k|^2} \cdot \vec \sigma + O(\delta k^2)\ .
\end{gather}
Enforcing a normalization of $\rho_k$ amounts to simultaneously
rescaling $a_k$ and $\wt{\vec b}_k$, under which the result is
invariant. Upon completing a Wilson loop, one obtains
\begin{gather}
  \label{2band-general}
  U_{0, 2\pi} = \mathcal{P}\exp \left[-i\int\limits_0^{2\pi} d \vec{b}_k \times \vec{b}_k \cdot \vec \sigma\right]\quad , \quad
  \vec{b}_k \equiv {\wt{\vec b}_k\over\sqrt{a_k^2 + \wt b_k^2}}
\end{gather}
where $\mathcal{P}$ denotes path ordering. 

\subsection{Projector limit}
In the limit $a_k = |\wt{\vec b}_k| = \frac{1}{2}$, e.g. at zero
temperature, $\{\rho_k\}$ are projectors, and one has
\begin{gather}
  U\nd_{0,2\pi} = \mathcal{P}\exp\!\left[-\frac{i}{2}\int\limits_{k=0}^{2\pi} d\hat{\vec b}_k
  \times \hat{\vec b}_k \cdot \vec \sigma\right]\ ,
\end{gather}
where $\hat{\vec b}_k = \vec b_k/|b_k|$.

\subsection{Coplanar limit}
An interesting limit is when the vectors $\vec b_k$ for different $k$
are restricted to a plane which \emph{contains the origin}. Examples
include the SSH model, graphene, or in the Haldane model where two
edge spectra become degenerate. In this limit one can drop the path
ordering and get
\begin{gather}
  U\nd_{0, 2\pi} = \exp\big(i\,2S\, \sigma_{\perp}\big)\quad , \quad S =
  \frac{1}{2}\,\hat {\vec n}_{\perp} \cdot \oint {\vec b}_k \times d{\vec
    b}_k
\end{gather}
where $S$ is the directed area spanned by the path of $\vec{b}_k$ and
$\sigma_{\perp}$ is the Pauli matrix along the normal direction of the
plane. In the projector limit, ${\vec b}_k$ lies on a circle of radius
$\frac{1}{\sqrt{2}}$ ({\it cf.\/} Eq. \ref{2band-general}), thus
$U\nd_{k, k+2\pi} = \exp(i n \pi)$ where $n$ is the winding number of
$\hat{\vec b}_k$.

Rotate to the frame where $\vec b_k$ is in the $(x,y)$ plane. Then an
unnormalized $\rho_k$ has the form $\rho_k =
\frac{1}{2}\Tr(\rho_k)\left(
  \begin{smallmatrix}
    1 & m_k^{*}\\ m_k & 1
  \end{smallmatrix}\right)$ where
\begin{gather}
  |m| = \frac{f\nd_> - f\nd_<}{f\nd_> + f\nd_<}
\end{gather}
with $f\nd_>$ and $f\nd_<$ the two eigenvalues of $\rho_0$. In the
rotated basis, $U\nd_{0, 2\pi} = \textsf{diag}\big( e^{2iS}\,,\, e^{-2iS}\big)$.
Thus $M = \rho\nd_0\, U\nd_{0,2\pi}$ has eigenvalues
\begin{gather}
  m^{\pm} = \frac{1}{2}\Tr(\rho_0) \left[ \cos(2S) \pm \sqrt{|m|^2 -
      \sin^2(2S)} \right]\ .
\end{gather}

\bibliographystyle{apsrev-no-url}
\bibliography{uhlmann}